# Governance in Adaptive Normative Multiagent Systems for the Internet of Smart Things: Challenges and Future Directions

Marx Viana[1], Lauro Caetano[1], Francisco Cunha[1], Paulo Alencar[2], Carlos Lucena[1]

[1]*Laboratory of Software Engineering (LES), Pontifical Catholic University-PUC-Rio*
{mleles, lcaetano, fcunha, lucena}@inf.puc-rio.br

[2]*University of Waterloo – Waterloo, Ontario – Canada*
palencar@uwaterloo.ca

**Abstract**

*The rapidly changing environments in which companies operate to support the Internet of Things (IoT) and Autonomous Vehicles is challenging traditional Multi-agent System (MAS) approaches. The requirements of these highly dynamic environments gave rise to Adaptive Normative MAS approaches. At the same time, governance is an essential and challenging feature that still needs to be addressed in adaptive normative MAS. Indeed, governance of individual and societal agent behavior in Adaptive Normative MASs is still a vague concept that has not been properly investigated, modeled and implemented. However, governance is fundamental for solving problems involving MAS coordination, organizations and institutions. In this paper, we present our ongoing research towards understanding and improving governance in Adaptive Normative MASs. We also discuss challenges and future directions that will facilitate the development of domain-specific smart IoT systems with governance features.*

## 1. Introduction

Multi-agent systems are often understood as complex entities where a multitude of agents interact, usually with some intended individual or collective purpose [20]. Such a view usually assumes some form of structure, or set of norms or conventions that articulate or restrain interactions in order to make them more effective in attaining those goals, more certain for participants or more predictable. The engineering of effective regulatory mechanisms is a key problem for the design of open complex multi-agent systems.

In recent years, coordination, organizations, institutions, and norms are four key governance elements and they have become a major issue in MAS research [16]. Moreover, recent applications of MAS on Internet of Smart Things, Autonomous Vehicles, Web Services, Grid Computing and Ubiquitous Computing enforce the need for using these aspects in order to ensure social order within these environments. Openness, heterogeneity, and scalability of MAS pose new demands on traditional MAS interaction models. The view of coordination and control has to be expanded to consider not only an agent-centric perspective but also MAS societal and organization-centric views as well.

Thus, the use of agents for construction of such complex systems is considered a promising approach in many areas [21]. However, the successful and widespread deployment of large-scale MASs requires a unifying set of agent related abstractions to support (meta-)modeling languages and their respective methodologies. Furthermore, there is still a poor understanding about some MAS abstractions such as those used to define different forms of adaptation and to represent concepts involving governance behavior [10]. In addition, most modeling languages do not define how these abstractions interact at runtime [2], but many software applications need to change their behavior and react to changes in their environments dynamically.

In general, modeling languages should represent the dynamic and structural aspects of software agents, expressing the essential characteristics of the entities. Structural aspects incorporate the definition of entities, their properties and their relationships. Several authors recognize the importance of modeling the agents in their environments both at design time and at runtime [7] [2]. These authors also observed that most (meta-)modeling languages do not represent some important concepts present in SMAs, such as adaptation and norms [2] [11]. These concepts were treated in [19] and an experimental evaluation demonstrated the validity of the study.

However, these concepts, when applied in the context of Internet of Smart Things, did not prove sufficient to solve governance problems in autonomous and complex systems. To deal with this challenge it is necessary to add to the metamodeling language proposed in [19] the concepts of coordination, organization, institution to support governance features. Without modeling these new features, it will not be possible to represent certain classes of relevant and naturally solvable problems by SMAs, such as those related to autonomous vehicles [14], smart lights [8] [17], smart cities [6]. In contrast, modeling these abstractions would make it possible to represent every smart thing that participates in an open environment as part of a governance oriented adaptive normative MAS, with sensors, actuators and algorithms capable of (autonomous) decision making as the environment undergoes changes.

In this context, we present our ongoing research towards understanding and improving governance in Adaptive Normative MASs. We also discuss challenges and future directions that will facilitate the development of domain-specific smart IoT systems with governance features. Our discussion will be based on a case study we have developed that focuses on intelligent intersections to prevent accidents. This case study showed that autonomous vehicles are promising for conflict resolution on intelligent intersections, assuming that the vehicles communicate with other. Currently, autonomous vehicles are often unable to negotiate their way into and out of busy intersections. This is a great challenge both in academy and industry.

This paper is organized as follows. Section 2 presents the related work. Section 3 describes and discusses a case study involving intelligent intersections. Section 4 discusses challenges and future directions of IoT-based MAS governance. Section 5 presents our conclusions.

## 2. Related Work

There are some research results in the literature about multi-agent systems that use specific types of coordination and governance approaches [1] [3] [5]. Chater et al., for example, provide a test case for computational theories of social interaction, with fundamental implications for the development of autonomous vehicles. The authors suggest that the challenge of understanding and building agents that can genuinely "negotiate" the traffic should be a major focus of cognitive science research. They state that this challenge is comparable in scale to the major research efforts in computer vision and machine learning that have focused on autonomous driving.

In [1], the authors propose a simulation framework to study the governability of complex socio-technical systems experimentally by means of agent-based modelling (ABM). This framework is rooted in a sociological micro-macro model of a socio-technical system, and considers the interplay of agents' choices (micro) and situational constraints (macro).

The authors in [3] claim that the multi-level governance concept encompasses vertical intra governmental relations and a more comprehensive process involving formal government agencies is needed. Multi-level governance is a challenging task and a requirement that is essential in the development of case studies involving intra governmental features.

In [17], the author proposes a governance and coordination mechanism based on communication to improve the management of traffic flows on intersections. Leadership is handed over among entities for a period of time and, as a leader is elected within the pool of cars, it becomes responsible for assigning traffic phases to incoming vehicles from all directions. In addition, the author suggests that the application features a safety alert when there is a traffic violation. However, no further mechanism of cognition or adaptation is provided to deal with normative behaviors.

All these solutions focus on governance systems, autonomous vehicles, smart things and multi-agent systems in isolation, but none of these approaches aims at creating a (meta-)model language to enhance the interaction between the users and the system, and introduces governance in societies of smart IoT things.

## 3. Motivating Case Study

To justify the relevance of governance-oriented adaptive normative MAS research for the Smart IoT, this section presents a case study involving intelligent intersections. Traditionally, the vehicle only reacts to the driver's commands. Recent advances in communications, control and embedded systems have changed this model, paving the way to autonomous vehicles. The car is now a extraordinary sensor platform, sensing information from the environment and the other cars, and feeding it to drivers and the infrastructure to assist in safe navigation and traffic management. The next step in this evolution is the intelligent intersection.

Our case study relates to vehicle normative behavior regulation around traffic intersections in Brazil. The number of cars is continuously growing in Brazil. The large increase in the Brazilian fleet brought the number of cars to one car for every 4.4 inhabitants, i.e., it is estimated that there are approximately 45.4 million private vehicles in Brazil. Ten years ago, the proportion was 7.4 inhabitants per vehicle [9]. With the increase in the number of vehicles on the streets and the arrival of autonomous vehicles, the need arose to create systems capable of assisting both traffic experts as well as self-driving agents to better deal with unexpected situations in day-to-day traffic. The right of way norms at traffic intersections, for example, are difficult to follow at intersections where there are no signs. Therefore, there are serious consequences when those norms are violated. An intersection is a junction where two or more roads meet, or cross.

According to data from the Brazilian Federal Highway Patrol [9], the main causes of fatal accidents in 2016 were, among others: lack of attention (30.8%); high speed (21.9%); alcohol consumption (15.6%); disregard for signs (10%); reckless overtaking (9.3%); and sleep (6.7%). In addition, 60% of these car accidents have occurred at intersections without signs. According to the Brazilian Transit Code (BTC), Article 29, the right of way norms for vehicles arriving at an intersection with no signs, are: (i) Norm1: vehicles moving on main roads have the preference; (ii) Norm2: in the case of a traffic circle, the one that is circulating through it has the preference, and (iii) Norm3: in other cases, vehicles coming from the right have the preference. In addition, Article 38, states that — before the driver makes a right or left turn onto other roads, the driver must, as per its Sole paragraph, yield to pedestrians, cyclists and vehicles that come from the opposite direction on the road they will be leaving, always respecting the norms of preference described in article 29.

The autonomous vehicles scenario consists of drivers, highways, traffic circles, and traffic intersections, respectively. The vehicles autonomous goal is to ensure that the smallest number of accidents happens. To achieve this goal, the driver agent must be restricted by norms, but due to its automation, it may decide to fulfill, or not, these norms. Such simulations are, in fact, normative multi-agent systems that receive data with the following information: (i) different traffic intersections, (ii) vehicle autonomous agents, (iii) norms to be followed by the vehicle autonomous agents, and (iv) different traffic scenarios. Simulations allow autonomous vehicles to find different solutions in order to save the lives of passengers during accidents at intersections.

Figure 1 presents our scenario: three cars arrive at an intersection at the same time. The agents' goals are: (i) The PINK autonomous vehicle wants to go ahead on street 1 and cross street 2; (ii) The YELLOW autonomous vehicle wants to go ahead on street 2 and cross street 1, and (iii) The RED autonomous vehicle is on street 1 and wants to turn left onto street 2. However, there are no traffic signs and the agents need to be able to make decisions to avoid collision among the cars considering the Brazilian traffic norms.

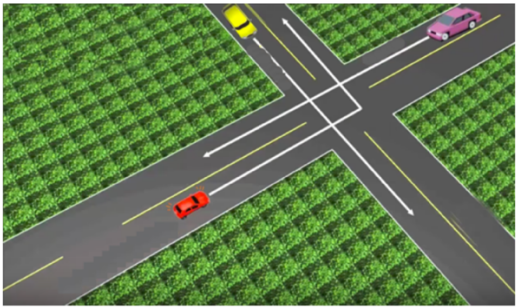

**Figure 1**: Traffic Intersection norms in Brazil.

As previously mentioned, articles 29 and 38 of the BTC deal with the norms of preference at intersections. However, neither Norm1 nor Norm2 of article 29 can be applied in this scenario. To solve this situation, the vehicle autonomous agents need to decide whether they will fulfill or violate Norm3 of article 29. This scenario considers that all autonomous vehicle agents fulfilled all the norms. The agents' internal reasoning was built by using the JSAN Framework [18] and they are described as if they had fulfilled Norm3: (i) the PINK autonomous vehicle agent arrived at the intersection and stopped because the YELLOW car is on its right; (ii) the YELLOW autonomous vehicle agent arrived at the intersection and stopped because the RED car is on its right, and (iii) the RED autonomous vehicle agent arrived at the intersection and there is no car on its right, therefore, the agent's reasoning cannot use article 29. To decide what to do, the agent needs to use article 38.

**Table 1** - Governance Strategies.

| Strategies | PINK | YELLOW | RED |
|---|---|---|---|
| *Pressured* | X | | |
| Rebellious | | X | |
| Social | | | X |

However, the Brazilian Transit Code (BTC) does not cover this situation, which creates a problem. As such, we need to improve a view of coordination and control has to be expanded to consider not only an agent-centric perspective but societal and organization-centric views as well. Sometimes, the analysis of the BTC articles mentioned above will not be enough to allow the agent to make a decision. Consequently, it is necessary to consider different types of governance-oriented adaptive normative MAS strategies that can be adopted by the normative agents, which will adapt their behaviors to deal with the norms.

For instance, in the scenario presented in Table 1: (i) the PINK autonomous vehicle agent adopts a pressured strategy, i.e., it fulfills the norms to achieve its individual goals, considering only the punishments that it will suffer; (ii) the YELLOW autonomous vehicle agent adopts a rebellious strategy, i.e., it considers only their individual goals and violates all of the environment's norms, and (iii) the RED autonomous vehicle agent adopts a social strategy, i.e., it complies with the norms and then verifies if it is possible to fulfill some of its individual goals. As a result, the PINK and RED autonomous vehicle agents give the preference to the YELLOW autonomous vehicle agent, which in turn accepts it because its rebellious strategy encourages this agent to go ahead.

## 4. Key Challenges and Future Directions

A major challenge to building governance-aware Adaptive Normative MASs results from the fact that autonomous multi-agent systems and of governance systems deal with a different set of abstractions, which have not yet been integrated [15] [4]. This barrier makes it difficult to integrate governance concepts with Internet of Smart Things based on MASs.

Several proposals have been provided in each of the areas, namely governance systems and smart things, namely investigations on areas related to governance research [3] [1] [12] [5] and on smart things based on multi-agent systems [8] [22] [13]. However, each of the respective research communities have worked in isolation, and we still lack a common and integrated solution that supports the integration of governance oriented adaptive normative MAS and smart things.

In response to the aforementioned main challenge, we are developing an approach, which involves a metamodel and an architecture, to introduce governance in adaptive normative software agents. Issues related to governance-oriented adaptive normative MAS and smart things include: (i) identify and analyze the limitations and the gaps of the different metamodels, architectures, languages and frameworks found in the literature related to modeling governance oriented in Adaptive Normative MASs; (ii) verify the feasibility of extending such metamodels, architectures, languages and frameworks, so that an infrastructure for the simulation of governance oriented in Adaptive Normative MASs is provided; (iii) create and propose a metamodel and an architecture to the simulation of governance in normative adaptive MASs for the development of smart things, such as smart traffic lights and autonomous vehicles; (iv) evaluate the metamodel and the architecture; (v) compare the approach with existing approaches, and (vi) develop use case scenarios and case studies to evaluate the proposed approach.

## 5. Conclusions

In this paper, we presented our ongoing research towards understanding and improving governance in Adaptive Normative MASs. We also discuss challenges and future directions that will facilitate the development of domain-specific smart IoT systems with governance features based on a case study we have developed that focuses on intelligent intersections to prevent accidents.

As future work, we will improve the governance in MASs by regulating the resources of the system through the proposed approach and ensuring that an agent is punished when it violates governance rules based on an evaluation that uses testimonies of other agents. Furthermore, we will explore domain-specific smart IoT domains that can benefit from governance features, which may include smart traffic lights and autonomous vehicles. Last, but not least, we will provide a metamodel, an architecture and an infrastructure to facilitate building governance oriented smart things based on adaptive normative MASs.